\documentclass[10pt,a4paper]{article}
\usepackage{amsmath,amsfonts,amssymb,bm,mathrsfs}
\usepackage{graphicx,color}
\usepackage{soul} 
\usepackage[utf8]{inputenc}
\usepackage[T1]{fontenc}
\usepackage{url}
\graphicspath{{./Figs/}}
\definecolor{MidRed}{rgb}{0.78,0,0}
\definecolor{MidGreen}{rgb}{0,0.65,0}
\definecolor{MidBlue}{rgb}{0,0,0.68}

\newcommand{\unit}[1]{\ensuremath{\mathrm{#1}}}

\newcommand{\req}[1]{(\ref{#1})}

\def\pageheads{C. E. Calosso, A. C. Cárdenas Olaya and E. Rubiola, \hfill~\today\hfill}
\title{Phase-Noise and Amplitude-Noise Measurement\\[0.5ex] of DACs and DDSs}
\author{Claudio E. Calosso$^\nabla$, A. Carolina Cárdenas Olaya$^\nabla$,\\[0.5ex]
and Enrico Rubiola$^{\exists\,\nabla}$\\[1em]
\normalsize\sffamily
$\nabla$ Istituto Nazionale di Ricerca Metrologica INRiM, Italy\\[1ex]
\normalsize\sffamily
$\exists$ FEMTO-ST Institute, UBFC and CNRS, France\\[1em]  
\large\sffamily\bfseries
http://rubiola.org}

\date{\normalsize\sffamily\today}

\begin{document}
\maketitle

\begin{abstract}
This article proposes a method for the measurement of Phase Noise (PN, or PM noise) and Amplitude Noise (AN, or AM noise) of Digital-to-Analog Converters (DAC) and Direct Digital Synthesizers (DDS) based on modulation-index amplification.  The carrier is first reduced by a controlled amount (30-40 dB) by adding a reference signal of nearly equal amplitude and opposite in phase.  Then, residual carrier and noise sidebands are amplified and sent to a conventional PN analyzer.  The main virtues of our method are: (i) the noise specs of the PN analyzer are relaxed by a factor equal to the carrier suppression ratio; and, (ii) the capability to measure the AN using a PN analyzer, with no need for the analyzer to feature AN measurement.  An obvious variant enables AN and PN measurements using an AN analyzer with no PN measurement capability.  Such instrument is extremely simple and easy to implement with a power-detector diode followed by a FFT analyzer.  Unlike the classical bridge (interferometric) method, there is no need for external line stretcher and variable attenuators because phase and amplitude control is implemented in the device under test.  In one case (AD9144), we could measure the noise over 10 decades of frequency.  The flicker noise matches the exact $1/f$ law with a maximum discrepancy of $\pm1$ dB over 7.5 decades.  Thanks to simplicity, reliability, and low background noise, this method has the potential to become the standard method for the AN and PN measurement of DACs and DDSs. 
\end{abstract}

\section{Introduction and State of the Art}
In virtually all domains of technology, the RF electronics is going digital via dedicated hardware, FPGA processing, Software Defined Radio techniques, and   ADCs and DACs are ubiquitous. 
This major trend is obviously driven by big telecom Companies, for mass consumer products 
and infrastructure equipment.
While basic principles of conversion are rather mature \cite[Chap.~1--3]{Kester-2004}, all the development is confidential.  
The technical information about converters and digital frequency synthesis is now in magazines \cite{Murmann-2015,Neu-2017,Cordesses-2004b,Cordesses-2004b,Verhelst-2015,Winoto-2018} and books \cite{Clara-2013}, \cite{Harpe-2015}, \cite[Chap.~3]{Hoefflinger-2016}, \cite{Manganaro-2012}, \cite[Chap.~ 9--11]{Manganaro-2013}, \cite{Symons-2013}.

Converters are available from leading manufacturers (chiefly, Analog Devices, Linear Technology, and Texas Instruments) with several GHz clock speed, 12-16 bits (BUS), and up to 12-13 ENOB (Equivalent Number Of Bits).  High-speed ADCs are generally more complex than DACs, and have inferior tradeoff between ENOB and maximum clock frequency.  The reason is that most ADC architectures (SAR, pipelined, and subranging flash) employ a DAC.

Our interest is oriented to scientific applications, where the demand for high-purity RF signals is ever growing.  The relevant parameters are low PM and AM noise, high stability, frequency agility, and programmable amplitude and phase. We have in mind general purpose instruments, Atomic-Molecular-Optics physics and atomic clocks \cite{Ryan-2017,Perego-2018,Francois-2015}, long-distance synchronization via fiber links \cite{Predehl-2012,Calonico-2014,Lopez-2012}, real-time phase measurements \cite{Calosso-2013-EFTF}, particle accelerators \cite{Accelerator-conferences}, etc.  

In this context, we focus on the AM and PM noise of DACs and DDSs.  
Interestingly, modern high-speed telecom-oriented DACs have an internal NCO, which makes the DAC very similar to the DDS.  If not, the NCO can be implemented in FPGA, transferring the data to the DAC via the JESD204B interface.  
Thus, we refer to the term DAC as a placeholder for both DAC and DDS.

Going through numerous data sheets, we see that manufacturers are most concerned with SFDR, SINAD, THD+R and ENOB, and leakage from/to adjacent channels  (see for example \cite[Chap.~2]{Kester-2004} for the definition of these  terms).  By contrast, phase noise is generally documented only as a typical plot of $\mathscr{L}(f)$ in a reference condition.  
It is often difficult to divide the device's phase noise from the contribution of the reference oscillator and of the PN analyzer.  
The literature says little or nothing about phase noise and about how it is measured, and nothing has been found about measurement methods specific to DACs.   The AM noise is neither seen in datas heets nor in technical literature.

We describe the noise as the PSD (Power Spectral Density) of the random phase $\varphi (t)$ and fractional amplitude $\alpha(t)$, denoted with $S_\varphi(f)$ and $S_\alpha(f)$ as a function of the Fourier frequency $f$.  Notice that the more popular quantity $\mathscr{L}(f)\triangleq\frac{1}{2}S_\varphi(f)$ is defined only for phase noise.  We use the polynomial law $S_\varphi(f)=\mathsf{b}_0+\mathsf{b}_{-1}/f+\dots $  and $S_\alpha(f)=\mathsf{h}_0+\mathsf{h}_{-1}/f+\dots$ truncated after the flicker term $1/f$, as appropriated for two-port components.  The reader may refer to \cite{IEEE-STD-1139-2008,Rutman-1991,Rutman-1978} for an introduction to PM noise, and to \cite{Rubiola-2005-AM} for AM noise.

Measuring some DDSs with various methods, we observed that the white noise coefficient $\mathsf{b}_0$ can be of $-165$ \unit{dBrad^2/Hz}.  The flicker coefficient $\mathsf{b}_{-1}$ can be of $-135$ \unit{dBrad^2} at 10~MHz output frequency, and of $-110$ \unit{dBrad^2} at 100-150~MHz output.  We published only a part of this at conference \cite{Calosso-2012}.  Anyway, these numerical values define the minimum requirement for the background noise.

\begin{figure}[t]\centering
\includegraphics[scale=0.648]{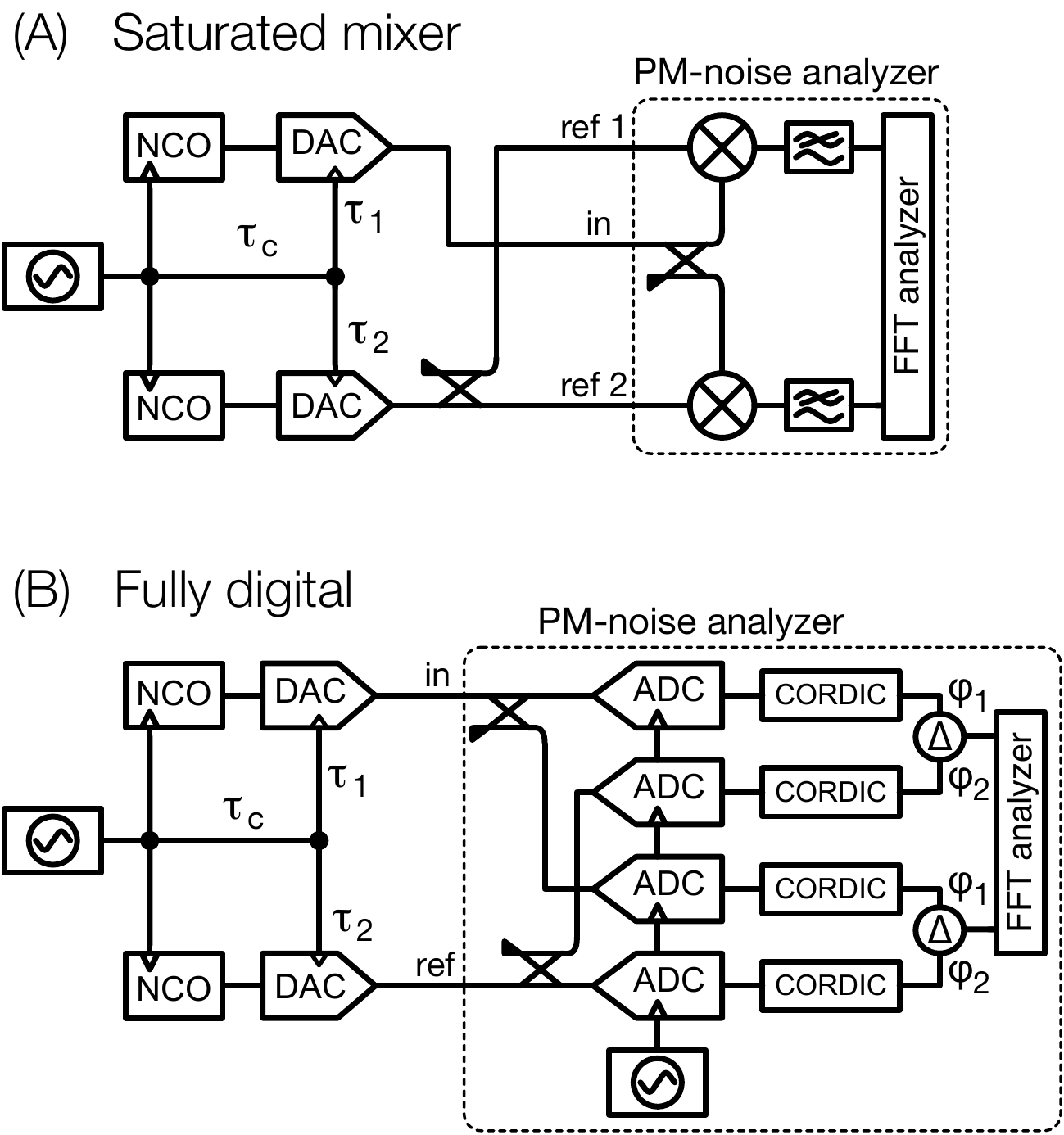}
\caption{Traditional phase-noise analyzers used for the measurement of DACs.  The PM noise spectrum results from the contribution of the two DACs.  
Notice that the single-DAC fluctuation $\tau_1$ and $\tau_2$ are detected, while  the clock fluctuation $\tau_c$ is rejected.}
\label{fig:Traditional-methods}
\end{figure}

Figure~\ref{fig:Traditional-methods} shows the direct measurement of the DAC noise with commercial PN analyzers.  All such instruments achieve reduced background noise by correlating and averaging the output of two channels.

The classical PN analyzer is based on a saturated mixer close to the quadrature condition, which converts the input phase into a voltage.  
In our case the quadrature condition can set numerically, provided the symmetry be sufficient to set the two channels at once.
A problem is the high saturation power of the mixer (7-15 dBm), compared to the low output power of the DACs ($\mathrm{\approx}$0 dBm).  The power splitters introduce additional 3 dB intrinsic loss.  The signal can be amplified, but amplifiers add complexity and noise.  

Conversely, the fully-digital analyzer is based on the direct AD conversion of the input signal \cite{Grove-2004}.  The benefit is obvious, in that the instrument accepts different frequencies at the `in' and `ref' inputs, and of course there is no phase adjustment.
At the time of writing there are only two options, Microsemi (formerly Symmetricom) \cite{Microsemi-ADEV} and, Jackson Labs \cite[scheduled mid/late 2019]{PhaseStation}.  
The Rohde Schwarz FSWP \cite{FSWP,Feldhaus-2016} is a digital instrument with down conversion from microwaves, but it has not a suitable `ref' input.  

Our method relies on modulation-index amplification by a factor $1/\eta\gg1$, introduced later.  After amplification, the AM and PM noise is so high that correlation is not necessary.  This solves two problems at once.  The first is that the correlation instruments rely on the hypothesis that the two channels are statistical independent.  This is sometimes untrue, and anyway hard to check.  Gross errors are around the corner if the experimentalist has not a deep understanding (see, for example \cite{Nelson-2014,Gruson-2017}).
The second problem is the measurement time, chiefly with the digital instruments because the noise of the DACs under test is often lower than that of the input ADCs.
In fact, the noise rejection is proportional to $1/\sqrt{m}$, where $m$ is the number of the FFTs averaged \cite{Rubiola-2010-xsp}.  This means 5 dB per factor of 10.  Accordingly, if a FFT starting from 1 mHz takes 2000 s acquisition time, averaging over 100 spectra for 10 dB noise rejection takes a measurement time of 2 days and 7.5 hours.  By contrast, our method needs a small $m$, only to smooth the measured spectra.

\section{Principles and Method}
\begin{figure}\centering
\includegraphics[scale=0.648]{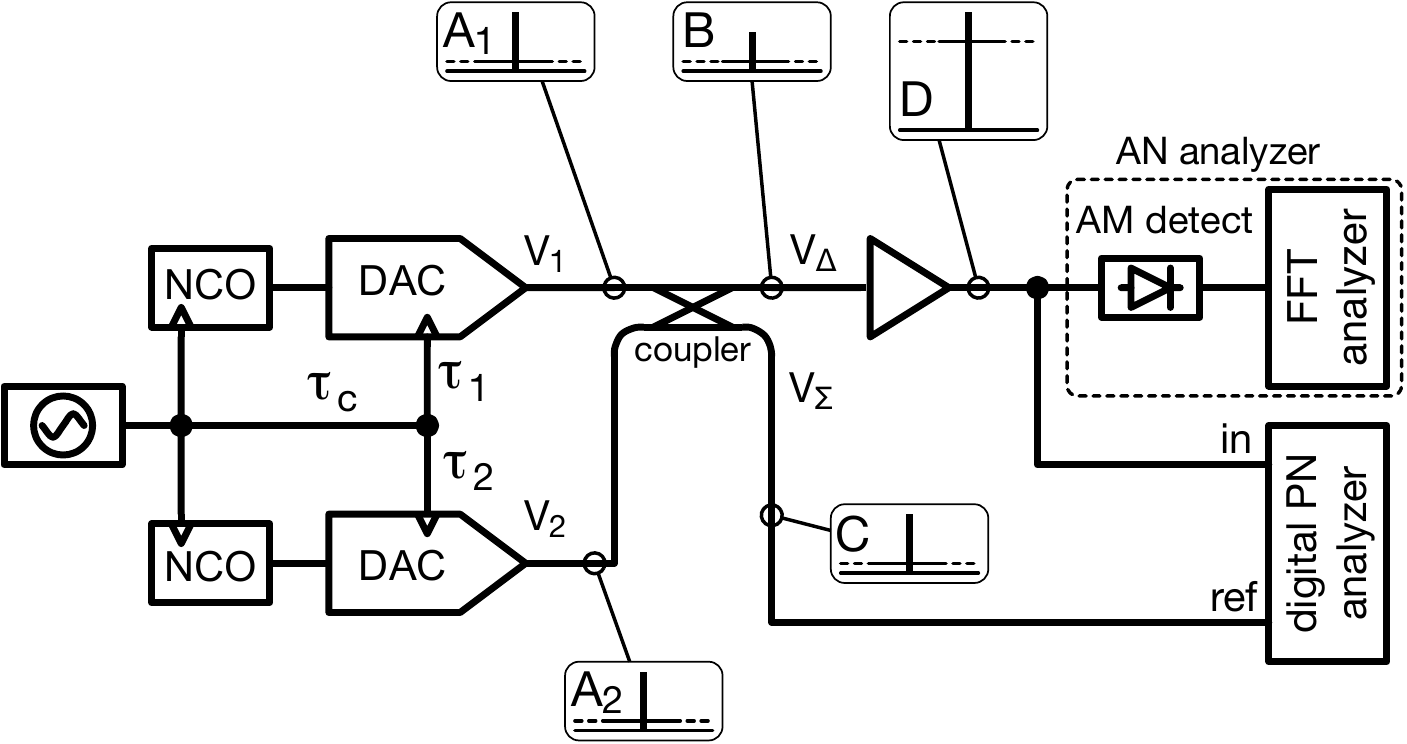}
\caption{Measurement method.  The rounded rectangles show the relevant spectra in log scale.}
\label{fig:Method}
\end{figure}

We compare two equal devices under test (DUT) using the scheme of Figure 2.  The rectangles labeled A${}_{1}$, A${}_{2}$, B, C and D show the spectrum in the critical points, in log scale.  The two signals $V_1$ and $V_2$ are combined in the directional coupler so that almost all the carrier power goes at the $\mathrm{\Sigma }$ output, and a small power goes to the $\mathrm{\Delta }$ output.  However, the noise sidebands are equally split between the $\mathrm{\Sigma }$ and $\mathrm{\Delta }$ output because these signals are not coherent (spectra B and C).  Residual carrier and noise sidebands are amplified (spectrum D) and detected by the PN analyzer or by the AN analyzer.

Our method derives from an early idea for the PN measurement of microwave two-port devices \cite{Sann-1968}.  In the original version, the reference carrier crosses the DUT, which contributes its own noise.  The carrier is completely suppressed by subtracting a copy of the reference through a magic Tee (a waveguide directional coupler), and the DUT noise sidebands are down-converted to DC by synchronous detection.  Microwave amplification of the noise sidebands was added later to boost the gain and to reduce the background noise \cite{Labaar-1982}.  The whole system is equivalent to a Wheatstone bridge powered by an AC signal (the microwave carrier), followed by a lock-in amplifier.  The lowest background noise is achieved after adjusting the system for the maximum carrier rejection \cite{Ivanov-1998-uffc,Rubiola-1999-RSI}.  By contrast, we leave a carefully controlled amount of carrier, as proposed in \cite{Walls-1997-ifcs} for different purposes.  There results modulation-index amplification, with optional AM-to-PM and PM-to-AM conversion depending on phase relationships.  The presence of the residual carrier is essential in that the signal is suitable to the measurement with all-digital PN analyzers, or with a simple AN analyzer implemented with a power-detector diode (\$100-300 for a packaged and connectorized detector) and a FFT spectrum analyzer \cite{Rubiola-2005-AM}.  Thanks to the AM-to-PM and PM-to-AM conversion, a PN analyzer with no AN measurement capability enables the measurement of both AN and PN.  Likewise, an AN analyzer with no PN measurement capability.  The modulation-index amplification relaxes the noise specs of the PN analyzer, or of the AN analyzer, by the same factor.  The RF gain (40-60 dB) is needed for the power to match the input range of the PN or AN analyzer.

The background white noise is due the noise $FkT$ of the amplifier, where $F$ is the amplifier noise figure (1-2 dB, that is, 1.25 to 1.6), and $kT\approx 4{\times}10^{-21}$ W/Hz is the thermal energy at room temperature.  Converted into PM noise, the background is $\mathsf{b}_0=2FkT/P$, where $P$ is the carrier power at the directional-coupler input, and the factor `2' accounts for the inherent 3-dB loss in the coupler.  The background $1/f$ PM noise is that of the amplifier, divided by the carrier suppression ratio.  This happens because the $1/f$ phase noise in components results from parametric up-conversion of the near-dc noise \cite{Boudot-2012}.

\subsection{Modulation-Index Amplification}
The directional coupler in Figure~\ref{fig:Method} delivers the output
\begin{equation}
\label{eq:VDelta}
V_\Delta = \frac{1}{\sqrt{2}}\left(V_2-V_1\right)~,
\end{equation}
where $V_1$ and $V_2$ are the output signals of the two converters, and the factor $1/\sqrt{2}$ is due to energy conservation in the absence of loss.  These signals have random fractional amplitude $\alpha_1(t)$ and $\alpha_2(t)$, and random phase $\varphi_1(t)$ and $\varphi_2(t)$.  We assume that  $\left|\alpha_1\right|\ll 1$, $\left|\alpha_2\right|\ll 1$, $\left|\varphi_1\right|\ll 1$ and $\left|\varphi_2\right|\ll 1$, hence $e^{\alpha +j\varphi}\simeq 1+\alpha +j\varphi$. 

For our purposes, it is convenient to set $V_1$ and $V_2$ close to the nominal amplitude $V_0$, but for a small difference in amplitude ($\beta_1$ and $\beta_2$) and phase ($\gamma_1$ and $\gamma_2)$, so that
\begin{align}
\label{eq:V1}
V_1 &= V_0 \left(1-\beta_1\right)e^{-i\gamma_1} e^{\alpha_1+j\varphi_1}\\
\label{eq:V2}
V_2 & = V_0\left(1+\beta_2\right)e^{+i\gamma_2} e^{\alpha_2+j\varphi_2}~.
\end{align}

\begin{figure}[t]\centering
\includegraphics[scale=0.648]{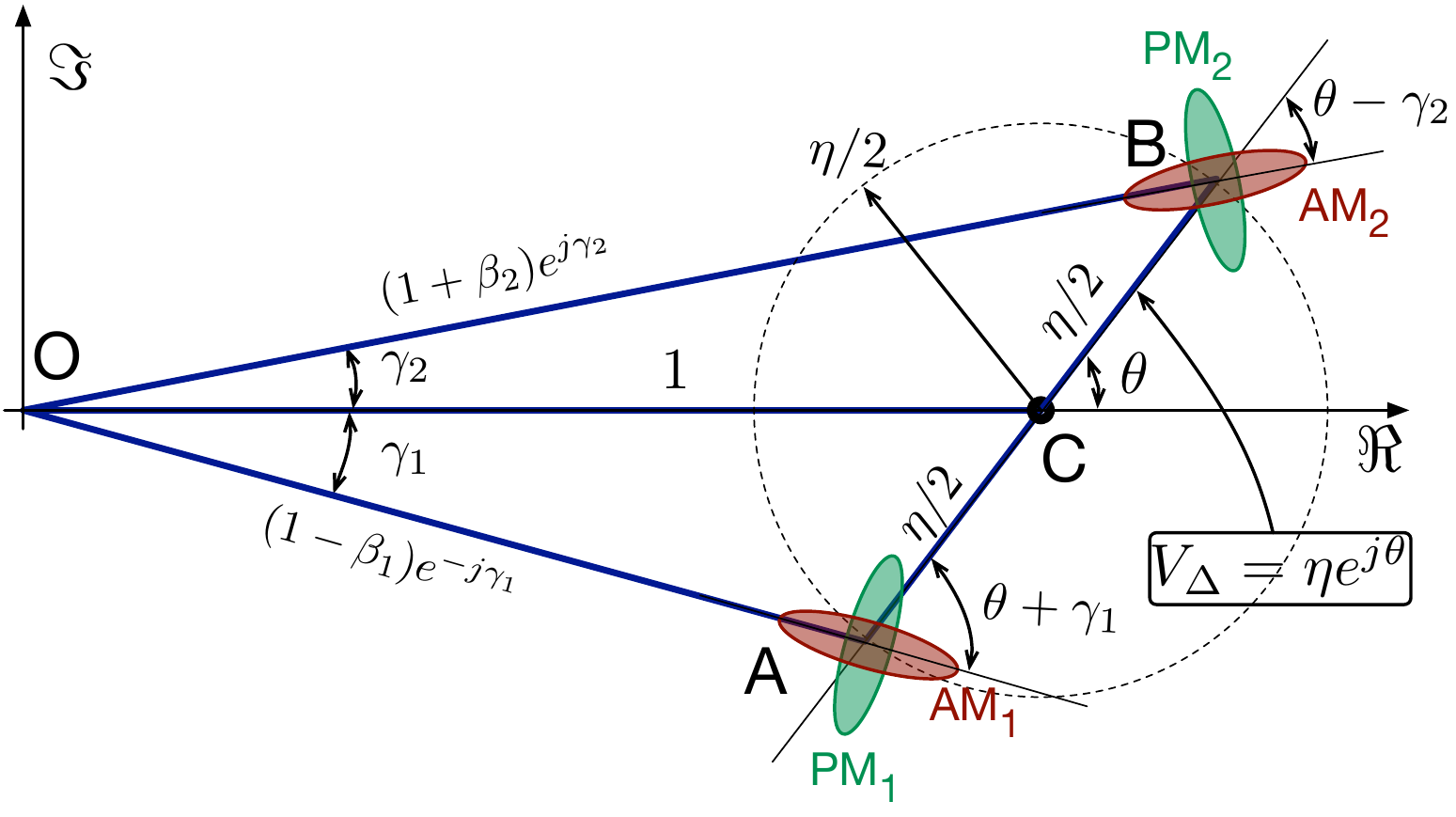}
\caption{Interplay between signals at the output of the directional coupler.  In the actual RF circuit, all signals are scaled up by a factor $V_0/\sqrt{2}$, omitted for graphical clarity.}
\label{fig:Modulation-amplification}
\end{figure}

\noindent
Figure~\ref{fig:Modulation-amplification} shows these signals and how they combine to form $V_\Delta$.  The latter can be written as
\begin{equation}
\label{eq:VDelta-complex}
V_\Delta = \eta \frac{V_0}{\sqrt{2}}\ e^{j\theta} e^{\epsilon +j\psi}~,
\end{equation}
where $\eta$ and $\theta$ describe the static amplitude and phase of the carrier seen at the coupler output, and $\epsilon(t)$ and $\psi (t)$ describe the amplitude and phase modulation.  We will show that $1/\eta$ is the modulation-index amplification.  The quantity $\eta$ has to satisfy $\left|e^{\epsilon +j\psi}\right|\ll\eta\ll1$.  First, $\eta\ll1$ is necessary for the amplification to be useful.  Second, $\eta\gg\left|e^{\epsilon +j\psi}\right|$ is needed to prevent $V_\Delta$ from sweeping $0$.  If this happens, PM are no longer defined.   Because AM and PM noise are small, these conditions allow a large range for $\eta$.

The desired amount of residual carrier is obtained by setting $\beta_1$, $\beta_2$, $\gamma_1$ and $\gamma_2$ to appropriate values.  Notice that $\beta_1$ and $\beta_2$, and likewise $\gamma_1$ and $\gamma_2$, are not equal in the general case, but they can be assumed equal for $\eta\ll1$.  Accordingly, it is useful to set $\beta_1=\beta_2=\beta/2$ and $\gamma_1=\gamma_2=\gamma/2$, which defines $\beta$ and $\gamma$.  By inspection on Fig.~\ref{fig:Modulation-amplification}, it holds that $\beta\ll1$ and $\gamma\ll\theta$ for $\eta\ll1$.

Combining \req{eq:VDelta}-\req{eq:V1}-\req{eq:V2} gives
\begin{align}
V_\Delta &= 
\frac{V_0}{\sqrt{2}}\left[\left(1+\frac{\beta}{2}\right)e^{j\gamma/2}e^{\alpha_1+j\varphi_1}
-\left(1-\frac{\beta}{2}\right)e^{-j\gamma/2}e^{\alpha_1+j\varphi_1}\right]~.
\label{eq:VDelta-expanded}
\end{align}
The AM and PM associated to $V_\Delta$ result from 
\begin{equation}
\label{eq:epsilon-psi}
\epsilon = \Re\left\{\frac{\mathcal{S}}{\mathcal{C}}\right\} 
\qquad \text{and} \qquad
\psi =\Im\left\{\frac{\mathcal{S}}{\mathcal{C}}\right\}~,
\end{equation}
where
\begin{align}
\mathcal{S} &=
\frac{V_0}{\sqrt{2}}\left[\left(1+\frac{\beta}{2}+j\frac{\gamma}{2}\right)\left(\alpha_2+j\varphi_2\right)
- \left(1-\frac{\beta}{2}-j\frac{\gamma}{2}\right)\left(\alpha_1+j\varphi_1\right)\right]
\end{align}
is the voltage swing of \req{eq:VDelta-expanded}, and
\begin{equation}
\mathcal{C} =\eta\frac{V_0}{\sqrt{2}}e^{j\theta}
\end{equation}
is the carrier.  Recalling that $\beta\ll1$ and $\gamma\ll\theta $ for $\eta\ll1$, Equations \req{eq:epsilon-psi}
%
can be elegantly rewritten as 
\begin{equation}
\label{eq:rotation}
\begin{bmatrix}\epsilon\\\psi\end{bmatrix} = 
\frac{1}{\eta}
\begin{bmatrix}\cos\theta&\sin\theta\\-\sin\theta&\cos\theta\end{bmatrix} \:
\begin{bmatrix}\alpha_2-\alpha_1\\\varphi_2-\varphi_1\end{bmatrix}
\end{equation}
The matrix represents a rotation by $\theta$, which comes from the fact that $I$ and $Q$ (in-phase and quadrature voltage swing) is projected on $V_\Delta$.  The term $1/\eta$ means that $I$ and $Q$ are now referred to a carrier of amplitude $\eta\ll1$, which results in proportionally larger amplitude or phase swing.  Equation \req{eq:rotation} emphasizes two key concepts
\begin{enumerate}
\item  The modulation index is amplified by a factor $1/\eta$
\item  The rotation enables to preserve the character of AM and PM ($\theta=0$), to interchange AM and PM ($\theta=\pi/2$), or to take any combination of AM and PM. 
\end{enumerate}
Replacing $\epsilon$ and $\psi$ with their spectra, \req{eq:rotation} becomes
\begin{equation}
\label{eq:rotation-spectra}
\begin{bmatrix}S_\epsilon\\S_\psi\end{bmatrix} = 
\frac{1}{\eta^2}
\begin{bmatrix}\cos^2\theta&\sin^2\theta\\\sin^2\theta&\cos^2\theta\end{bmatrix} \:
\begin{bmatrix}S_{\alpha2}+S_{\alpha1}\\S_{\varphi2}+S_{\varphi1}\end{bmatrix}~,
\end{equation}
which enables the measurement options listed in Table~\ref{tab:Options}.

It is important to remember \req{eq:rotation-spectra} relies on the assumption that the noise of the two converters is uncorrelated because the system is insensitive to common-mode noise.  This is  clear with the PM noise: the common-mode time fluctuation $\tau_c$ is rejected, which includes the clock (Fig.~\ref{fig:Method}).  By contrast, correlation in the AM noise is more subtle because it originates from the power supply and from the voltage reference.  Insufficient PSRR (Power-Supply Rejection Ratio) may result in correlated noise, and multiple DACs in a chip may share the reference.

\begin{table*}[t]\centering\normalsize
\caption{Measurement Options.}
\vspace{0.5ex}
\begin{tabular}{|c|c|c|c|} \hline 
\rule[-0.6em]{0.0ex}{1.8em}
\sffamily Operation & \sffamily PN-analyzer only & \sffamily AN-analyzer only& \sffamily Signal vectors \\ \hline 
\rule[-2em]{0.0ex}{4.5em}%
\parbox{6em}{\centering{\large$\theta =0$}\\[0.5ex]AN and PN\\amplification} 
& $S_\psi\simeq\dfrac{S_{\varphi2}+S_{\varphi1}}{\eta^2}$ 
& $S_\epsilon\simeq\dfrac{S_{\alpha2}+S_{\alpha1}}{\eta^2}$ 
& \raisebox{-1.2em}{\includegraphics[scale=0.354]{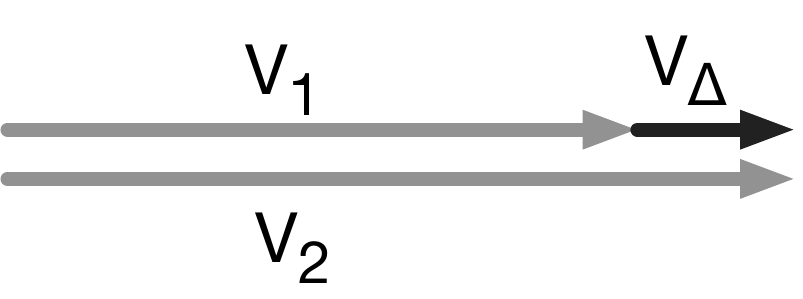}} \\ \hline 
\rule[-2em]{0.0ex}{4.5em}%
\parbox{6em}{\centering{\large$\theta=\pi/2$}\\[0.5ex]AN-PN cross\\amplification}
& $S_\psi\simeq\dfrac{S_{\alpha2}+S_{\alpha1}}{\eta^2}$ 
& $S_\epsilon\simeq\dfrac{S_{\varphi2}+S_{\varphi1}}{\eta^2}$ 
& \raisebox{-1.3em}{\includegraphics[scale=0.354]{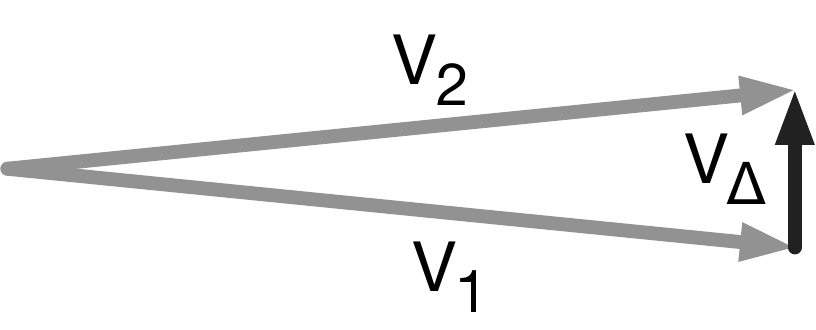}} \\ \hline 
\end{tabular}
\label{tab:Options}
\end{table*}

\subsection{Exact Analysis}
We evaluate the errors due to the hypothesis that $\eta \ll 1$.  With reference to Figure~\ref{fig:Modulation-amplification}, $\beta_1$ and $\beta_2$, and likewise $\gamma_1$ and $\gamma_2$ are not equal in the general case.  Combining \req{eq:VDelta}-\req{eq:V2} gives
\begin{align}
V_\Delta &= 
\frac{V_0}{\sqrt{2}}\Bigl[\left(1+\beta_2\right) e^{j\gamma_2}
e^{\alpha_1+j\varphi_1} 
- \left(1-\beta_1\right)e^{-j\gamma_1}e^{\alpha_1+j\varphi_1}\Bigr]~.
\end{align}
In low noise conditions, the associated voltage swing is 
\begin{align}
\mathcal{S} &=
\frac{V_0}{\sqrt{2}}\Bigl[\left(1+\beta_2\right)e^{j\gamma_2}\left(\alpha_2+j\varphi_2\right)
- \left(1-\beta_1\right)e^{-j\gamma_1}\left(\alpha_1+j\varphi_1\right)\Bigr]~.
\end{align}
Putting the above expression of $\mathcal{S}$ in \req{eq:epsilon-psi} gives
\begin{align}
\epsilon &=
\frac{1+\beta_2}{\eta}\Bigl[\alpha_2\cos(\theta-\gamma_2) +
\varphi_2\sin(\theta-\gamma_2)\Bigr] 
    +{}\nonumber\\&\qquad\qquad
-\frac{1-\beta_1}{\eta}\Bigl[\alpha_1\cos(\theta+\gamma_1) 
+\varphi_1\sin(\theta+\gamma_1)\Bigr]\\[2ex]
\psi &=
\frac{1+\beta_2}{\eta}\Bigl[-\alpha_2\cos(\theta-\gamma_2) +
\varphi_2\sin(\theta-\gamma_2)\Bigr] 
    +{}\nonumber\\&\qquad\qquad
-\frac{1-\beta_1}{\eta}\Bigl[-\alpha_1\cos(\theta+\gamma_1) +
\varphi_1\sin(\theta+\gamma_1)\Bigr]~.
\end{align}
Using the matrix expression of the above for the spectra of uncorrelated DACs, similar to \req{eq:rotation-spectra}, yields
\begin{align}
\begin{bmatrix} S_\epsilon\\S_\psi\end{bmatrix} &=
\frac{(1+\beta_2)^2}{\eta^2}\begin{bmatrix}A_2\end{bmatrix}\;\begin{bmatrix}S_{\alpha2}\\S_{\varphi2}\end{bmatrix} 
+ \frac{(1-\beta_1)^2}{\eta^2}\begin{bmatrix}A_1\end{bmatrix}\;\begin{bmatrix}S_{\alpha1}\\S_{\varphi1}\end{bmatrix}
\label{eq:matrix-exact}
\end{align}
with
\begin{align}
\begin{bmatrix}A_2\end{bmatrix} &=
\begin{bmatrix}
\cos^2(\theta-\gamma_2) & \sin^2(\theta-\gamma_2)\\
\sin^2(\theta-\gamma_2) & \cos^2(\theta-\gamma_2) 
\end{bmatrix}\\[1ex]
\begin{bmatrix}A_1\end{bmatrix} &=
\begin{bmatrix}
\cos^2(\theta+\gamma_1) & \sin^2(\theta+\gamma_1)\\
\sin^2(\theta+\gamma_1) & \cos^2(\theta+\gamma_1) 
\end{bmatrix}~.
\end{align}

We introduce the approximation that $S_{\alpha 1}\simeq S_{\alpha 2}$ and $S_{\varphi 1}\simeq S_{\varphi 2}$, based on the fact that the two DACs are nominally equal.  Thus, it makes sense to approximate \req{eq:matrix-exact} as
\begin{equation}
\label{eq:rotation-A}
\begin{bmatrix} S_\epsilon\\S_\psi\end{bmatrix} =
\frac{1}{\eta^2}\begin{bmatrix}A\end{bmatrix}\;
\begin{bmatrix}S_{\alpha2}+S_{\alpha1}\\S_{\varphi2}+S_{\varphi1}\end{bmatrix}~,
\end{equation}
averaging the two matrices
\begin{equation}
\label{eq:matrix-A}
\begin{bmatrix}A\end{bmatrix} =
\frac{1}{2}\left\{
\left(1+\beta_2\right)^2\begin{bmatrix}A_2\end{bmatrix} + 
\left(1-\beta_1\right)^2\begin{bmatrix}A_1\end{bmatrix}
\right\}~.
\end{equation}

\begin{figure}\centering
\includegraphics[scale=0.648
]{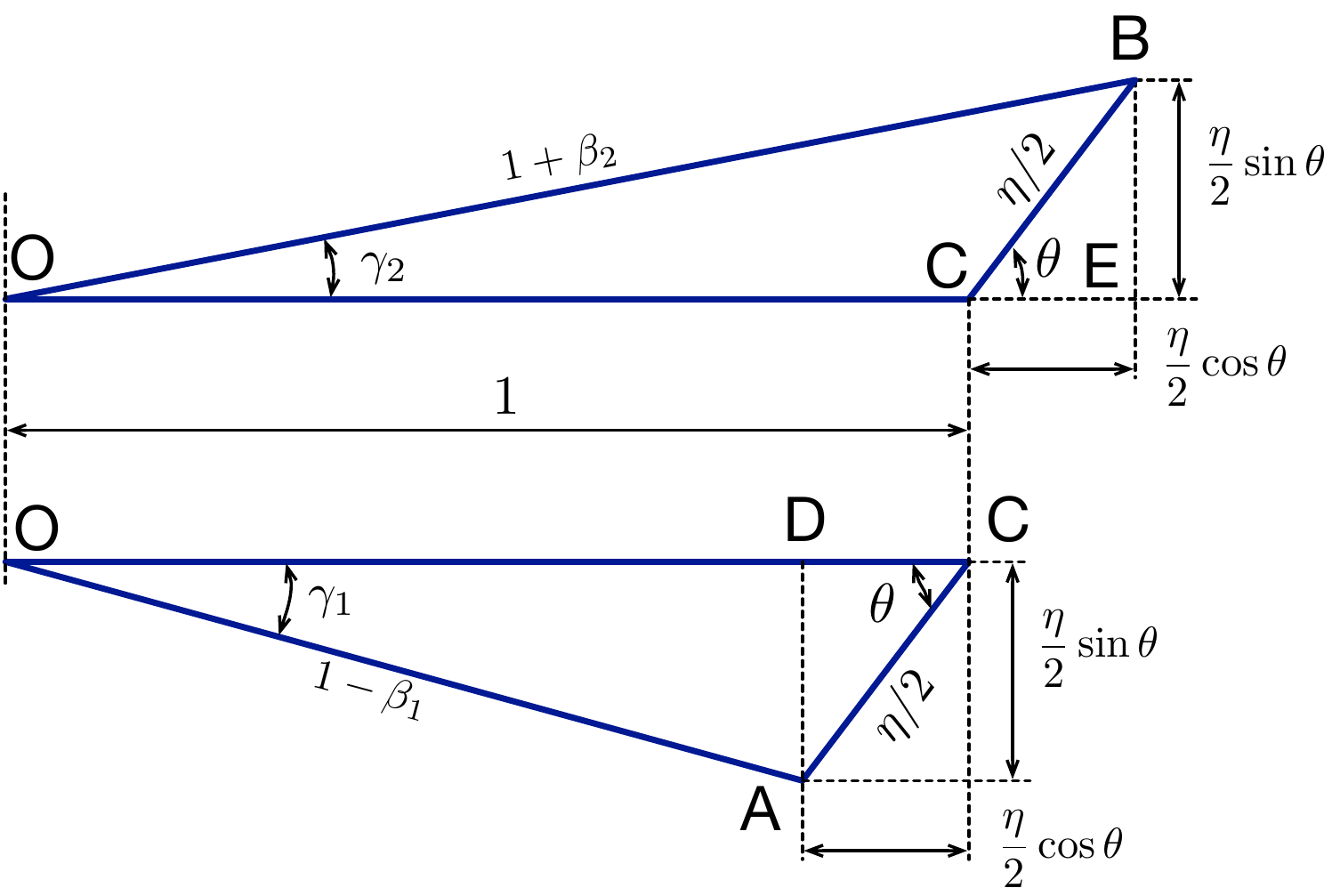}
\caption{Geometrical properties from Figure~\ref{fig:Modulation-amplification}.}
\label{fig:Geometry}
\end{figure}

\noindent From the geometrical properties shown on Figure 4, the Pythagoras theorem gives
\begin{align}
\label{eq:pb2}
\left(1+\beta_2\right)^2 &=
\left(1+\frac{\eta}{2}\cos\theta\right)^2 + 
\left(\frac{\eta}{2}\sin\theta\right)^2\\[1ex]
\label{eq:pb1}
\left(1-\beta_1\right)^2 &=
\left(1-\frac{\eta}{2}\cos\theta\right)^2 + 
\left(\frac{\eta}{2}\sin\theta\right)^2
\end{align}
\begin{align}
\label{eq:gamma2}
\gamma_2 &=
\arctan\left(\frac{\frac{\eta}{2}{\sin\theta}}{1+\frac{\eta}{2}{\cos\theta}}\right)\\[1ex]
\label{eq:gamma1}
\gamma_1 &=
\arctan\left(\frac{\frac{\eta}{2}{\sin\theta}}{1-\frac{\eta}{2}{\cos\theta}}\right)
\end{align}
Putting \req{eq:pb2}-\req{eq:gamma1} into \req{eq:matrix-exact} and expanding \req{eq:rotation-A}, we get a general expression for $\begin{bmatrix}A\end{bmatrix}$.  This expression is omitted here because it too complex and hard to read.  We discuss two relevant cases, derived the results with a symbolic math app.
The first case is 
\begin{equation}
\label{eq:lim-A}
\lim_{\eta\rightarrow0}\begin{bmatrix}A\end{bmatrix} =
\begin{bmatrix}
\cos^2(\theta) & \sin^2(\theta)\\
\sin^2(\theta) & \cos^2(\theta) 
\end{bmatrix}~,
\end{equation}
which is equivalent to \req{eq:rotation-spectra} and validates our hypotheses.

The second case is the accuracy of \req{eq:rotation-A} for $\theta=0$, still under the assumption that the noise of the two DACs is equal.  For this purpose, we allow a small angle error $\zeta$, and we expand \req{eq:matrix-exact} in series truncated to the second order of $\eta$ and $\zeta$.  For $\theta=0$, replacing $\theta\rightarrow\zeta$ gives
\begin{equation}
\label{eq:A-parallel}
\begin{bmatrix}A\end{bmatrix} = 
\begin{bmatrix}
1+\eta^2/4-\zeta^2 & \zeta^2\\[0.5ex]
\zeta^2 & 1+\eta^2/4-\zeta^2
\end{bmatrix}~.
\end{equation}
The third case is the accuracy of \req{eq:rotation-A} for $\theta=\pi/2$,  analyzed in the same way as before but for $\theta=\pi/2+\zeta$.  Hence,
\begin{equation}
\label{eq:A-orthogonal}
\begin{bmatrix}A\end{bmatrix} = 
\begin{bmatrix}
\zeta^2+\eta^2/4 & 1-\zeta^2\\[0.5ex]
1-\zeta^2 & \zeta^2+\eta^2/4 
\end{bmatrix}~.
\end{equation}
The finite gain results in small errors, negligible in most practical cases.  For example, $1/\eta^2=100$ (20 dB) results in $1.1{\times}10^{-2}$ dB error in \req{eq:A-parallel}, and in $-26$ dB coupling between AM and PM in \req{eq:A-orthogonal}.  
Of course, the gain must be well calibrated because of the $1/\eta^2$ factor in \req{eq:matrix-exact}.  The angle error deserves more attention.  For example, $\zeta=0.1$ ($5.7^\circ$) results in $-0.44$ dB error and $-20$ dB coupling between AM and PM in both \req{eq:A-parallel} and \req{eq:A-orthogonal}.
The PM noise is generally dominant because of the jitter in the clock distribution, and may pollute the AM-noise measurement.

\section{Experiments}

\begin{figure}[t]\centering
\includegraphics[scale=0.648,angle=90]{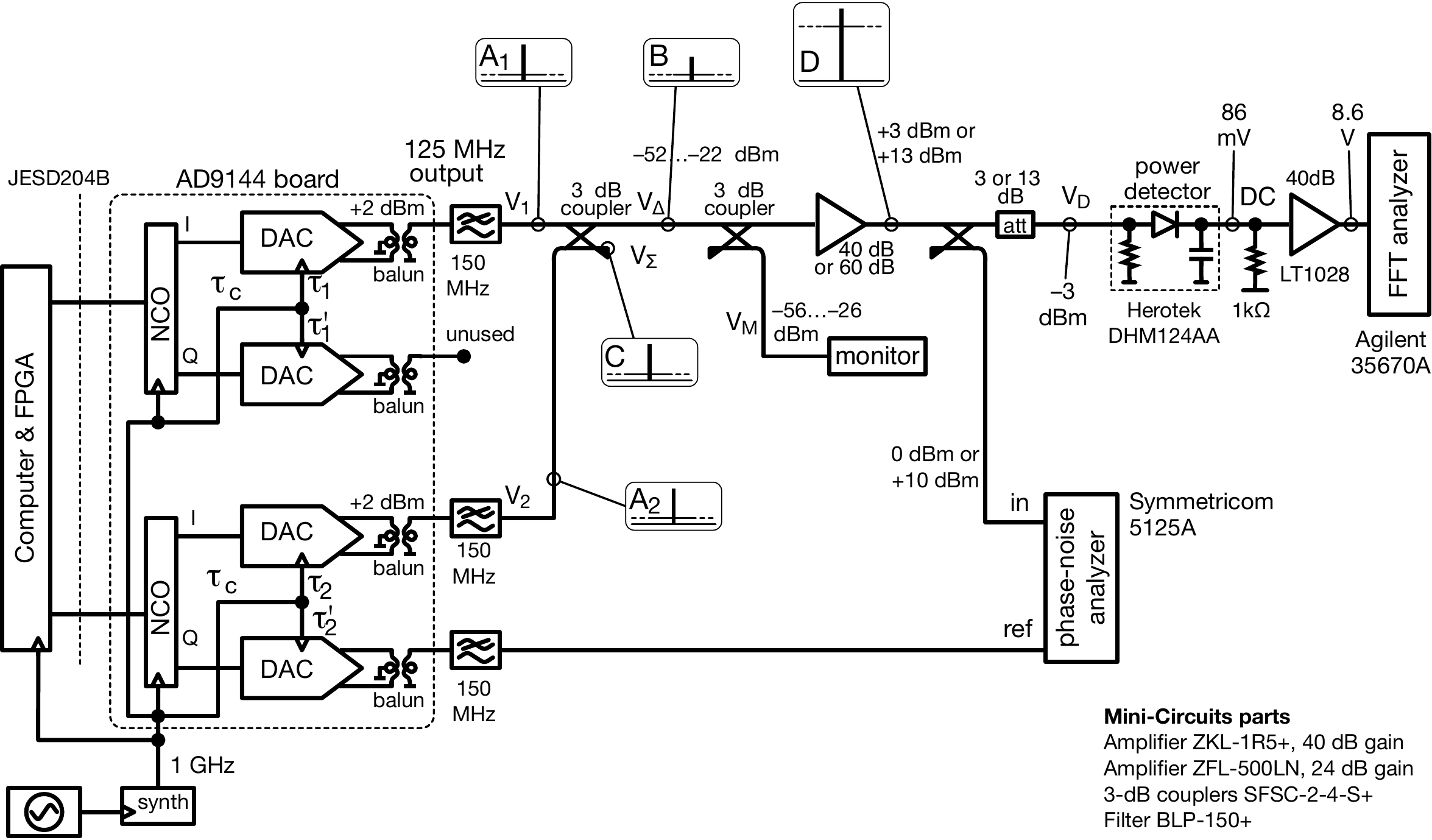}
\caption{Experimental setup.}
\label{fig:Block-diagram-full}
\end{figure}

\begin{table}
\def\vra{\rule[-0.6em]{0ex}{1.8em}}
\def\vrb{\rule[-0.4em]{0ex}{1.4em}}
\caption{Operating Parameters of Fig.~\ref{fig:Block-diagram-full}.}
\vspace{0.5ex}
\centering\normalsize
\begin{tabular}{|c|c|c|c|c|c|c|}\hline
\vra$\eta$&$V_\Delta$ & $V_M$ & Gain & $V_\text{in}$ & $V_D$ & $V_\text{DC}$\\\hline
\vrb$-20$ & $-22$ & $-26$ & $40$ & $+10$ & $-3$ & $86$\\\hline 
\vrb$-30$ & $-32$ & $-36$ & $40$ &  $0$  & $-3$ & $86$\\\hline
\vrb$-40$ & $-42$ & $-46$ & $60$ & $+10$ & $-3$ & $86$\\\hline
\vrb$-49$	& $-51$ & $-55$ & $60$ &  $+1$ & $-2$ & $86$\\\hline
\vrb  dB  &  dBm  &  dBm  &  dB  &  dBm  &  dBm &  mV\\\hline
\multicolumn{7}{|l|}{\vrb\small `$V$' given in dBm means `the power of the signal $V$.'}\\\hline
\end{tabular}
\end{table}

We measured the two channels of an AD9144 with the scheme of Fig.~\ref{fig:Block-diagram-full}.  The four channels of the AD9144 go in pairs, with two NCOs.  The two channels we measure can be considered as separated devices, but for a small crosstalk and for the small jitter $\tau_c$ in the common-mode clock path.  The AD9144 is driven by a Z-Board ZC706 via the JESD204B interface.  The ZC706 is a computer based on a Zynq-7000, with flash memory, DDR3 memory, Ethernet, USB, SD card slot, HDMI video, and SPI and FMC connectors for daughterboards.  It runs Linaro linux.  In turn, the Zync-7000 is a System on Chip (SoC), which is basically a microprocessor and a FPGA on the same chip.  

The AD9144 is actually an AD9144-FMC-EBZ card \cite{AD9144-card}, which contains the AD9144 chip, the AD9516-0 clock generator, the output baluns, and auxiliary functions.  This card is plugged onto the FMC connector of the Z-Board, and delivers $+2$ dBm power on 50 $\Omega$ load on each output (SMA connector).  The speed of the JESDB204B sufficient to set instantaneous phase and amplitude, but in our case such speed is needed only for the amplitude because phase and frequency are static parameters, set only once.  The sampling frequency is of 1~GHz, and the DAC output frequency is of 125 MHz.  This choice ($f_\text{ck}/f_\text{out}=8$, a small round number) eliminates the pseudo-random spurs, and makes it easy to measure the `true' random noise inherent in the component (see \cite{Nicholas-1987,Torosyan-2005} for the origin of such spurs).  The 1-GHz clock is generated by a Rohde Schwarz SMG 801 synthesizer driven by its internal low-noise OCXO.  The jitter of this clock is rejected because the clock is common mode.  There is no need to feed the PN analyzer directly with the OCXO signal because the PN analyzer has a reference input, derived from the DAC output. The noise of such reference has negligible effect because of the modulation-index amplification.  

Most of the tests were done in Torino with the exact scheme of Fig.~\ref{fig:Block-diagram-full}, and with $1/\eta$ is from 20 dB to 49 dB.  Table 2 shows the operating parameters.  However, preliminary tests were done in Besançon using a slightly different configuration.  We had less flexibility in the choice of the gain, and no power detector.  The output $V_\Sigma$ was used as the carrier-rejection monitor, and later in the experiment as the reference of the PN analyzer.  This saves one coupler on the RF path, and results in 3-dB lower background noise.

The AD9144 data sheet does not indicate the ENOB\@.  Instead, it gives the typical white noise $-162$ dBm/Hz ($N=6.3{\times}10^{-20}$ W/Hz) with 150 MHz single-tone output, full amplitude, and 983.04 MHz sampling frequency.  These conditions are quite similar to ours.  The white AN follows from $S_\alpha=N/P$, where $P$ is the output power.  The PN follows from $S_\varphi=N/P+J$, where $J$ is the additional contribution of the clock jitter, unknown here and not impacting on the AN\@.  From the specs, we expect $S_\alpha=-164$~dB/Hz with $+2$~dBm output power, and $S_\alpha=-161$~dB/Hz for the noise of two channels.  

Accounting for 7 dB loss from the DAC output to the input of the RF amplifier (low-pass filter and two 3-dB couplers), and adding the independent contribution of two DACs, the expected noise is of $-166$ dBm/Hz at the input of the RF amplifier.  This value is 8 dB higher than the thermal noise at room temperature ($-174$ dBm/Hz).  Allowing a noise figure of 2 dB, the amplifier contributes $+1$ dB to the noise limit.

The AN measurement is performed with a single-channel system (power detector, amplifier and FFT analyzer).
At 86~mV output, the power detector is no longer in the quadratic region, and detects the fluctuations of the RF peak voltage.  This operating mode is safe for regular AN measurement, but it would not be suitable to other applications, like beating two tones in the Pound frequency control.  
For reference, a flicker of $-100$ dB/Hz is equivalent to an amplitude stability of 11.8 ppm (use $\sigma^2=2\ln(2)\,\mathsf{h}_{-1}$ with $\mathsf{h}_1=10^{-100/10}$, thus $\sigma =1.18{\times}10^{-5}$).

\begin{figure}[t]\centering
\includegraphics[scale=0.648]{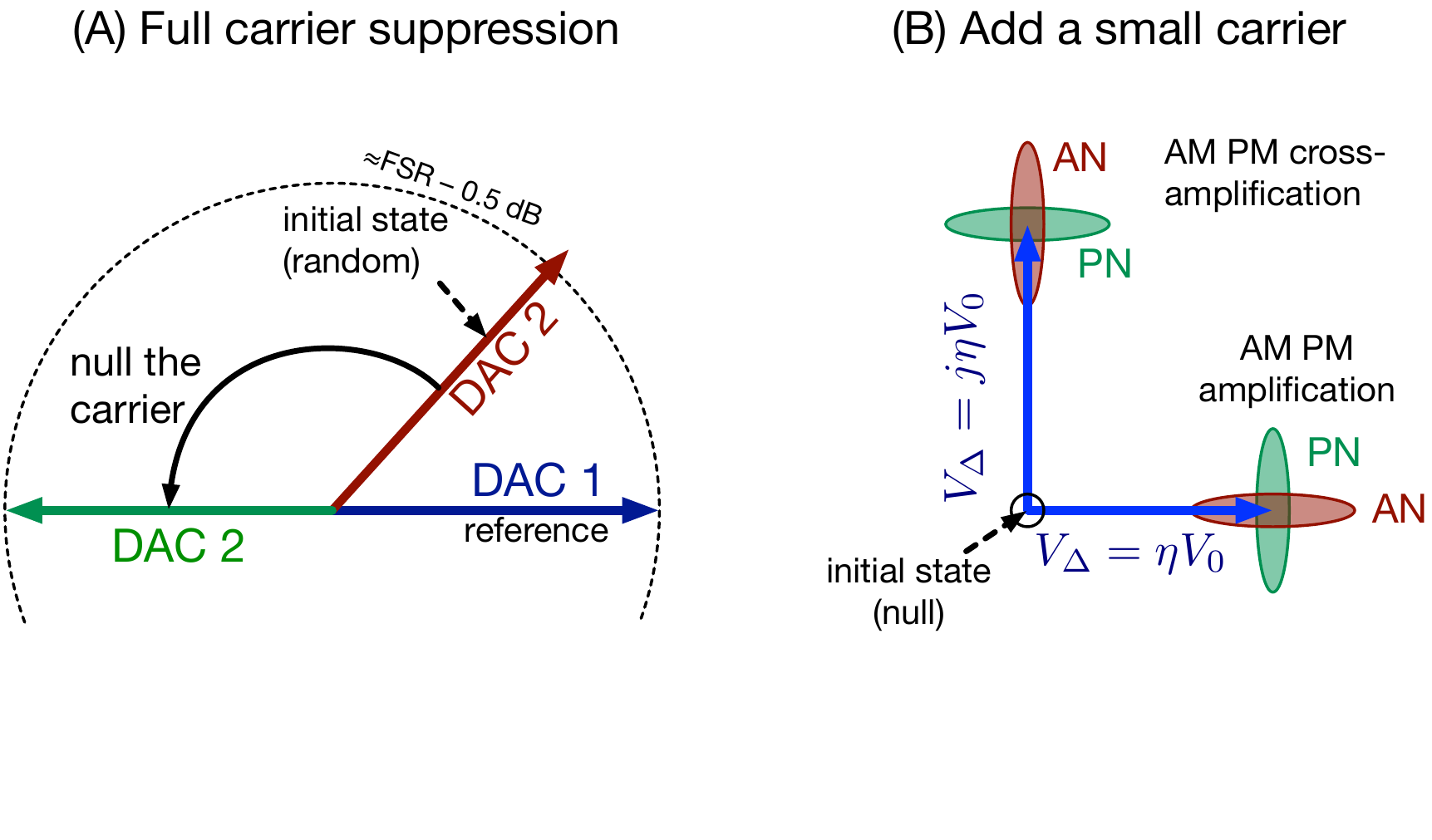}
\vspace{-18mm}
\caption{Adjustment procedure.}
\label{fig:Adjustment}
\end{figure} 

Before using, the system needs a simple adjustment and calibration.
We first need to set $\eta$ and $\theta$ to the desired value by adjusting phase and amplitude of $V_1$ and $V_2$.  With reference to Figure~\ref{fig:Adjustment}, we start with $V_1$ and $V_2$ just below the full-scale range.  In this condition, $V_1-V_2$ is determined by the arbitrary phase difference, plus the small amplitude asymmetry.  Then, we adjust phase and amplitude alternatively for maximum carrier suppression by monitoring the residual carrier.  The monitor is an oscilloscope triggered by the 125-MHz reference at one output of the AD9144, or a spectrum analyzer.  The spectrum analyzer is more comfortable to use, but the oscilloscope gives access to the sign of the error signal.  Setting phase and amplitude manually, we found that a sub-binary search is advantageous.  Albeit it takes more iterations than the `true' binary search, it is conservative vs uncertainty and noise in the monitor, and it simplifies the decision about the next iteration. 

After the maximum carrier rejection, we set the desired amount of residual carrier by changing $V_2$.  We recommend to check on AM-PM gain and crosstalk by adding a digital modulation to the DAC under test.

\section{Results}

\begin{figure}[t]\centering
\includegraphics[scale=0.385,angle=90]{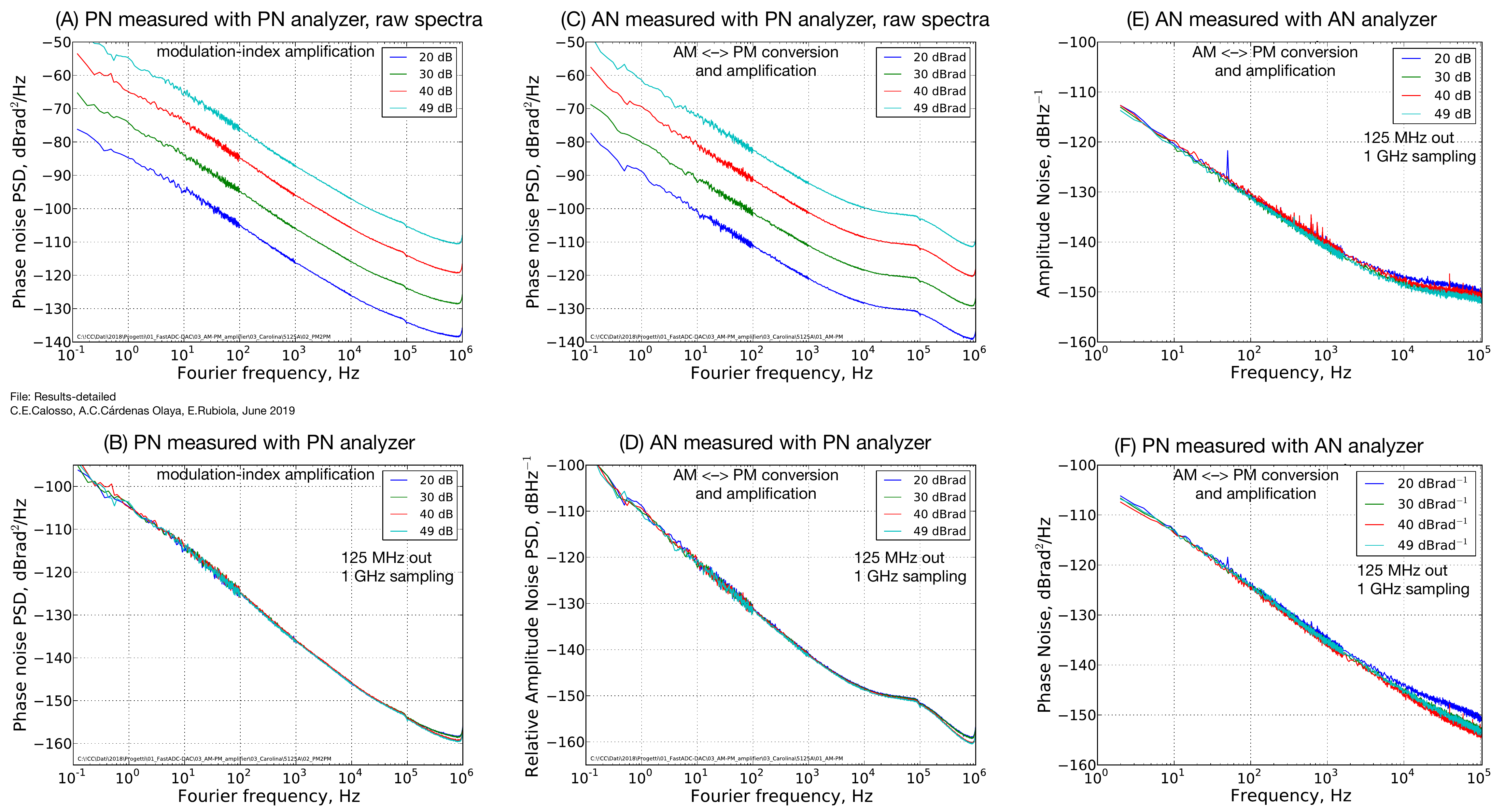}
\caption{Phase noise and amplitude noise of the AD9144, two channels.  The four options are shown, with modulation-index amplification and AM-PM cross amplification, and the two types of detection, AM and PM.  The spectra in (B) are the same of (A), but in (B) we account for the gain.  Likewise, (D) and (C).}
\label{fig:Results-detailed}
\end{figure}

Figure~\ref{fig:Results-detailed} shows the PM noise and AM noise of the AD9144 measured with different gain levels.  All the spectra in this Section refer to the total noise of the two output channels, hence the noise of one channel is 3 dB lower.  

In the first experiment (Fig.~\ref{fig:Results-detailed}\,A and \ref{fig:Results-detailed}\,B) we use the PN analyzer to measure both the PM and AM noise.  Figure \ref{fig:Results-detailed}\,A shows the raw phase noise spectra as displayed by the PN analyzer, for different values of the gain $1/\eta$ from 20 dB to 49 dB\@.  Figure~\ref{fig:Results-detailed}\,B reports the same spectra, corrected for the gain.  The plots overlap perfectly on almost all the frequency span, indicating that the value of $1/\eta$ is not critical.  In the upper half-decade, the noise is some 1 dB higher at lower gain.  We did not investigate further on this small discrepancy.  Likewise, Figure~\ref{fig:Results-detailed}\,C and \ref{fig:Results-detailed}\,D show the AM noise measured with the PN analyzer.  The residual carrier is orthogonal to the input carrier ($\theta=\pi/2$), so that the system performs AM-to-PM conversion. The raw spectra (Fig.~\ref{fig:Results-detailed}\,C) are correctly given in \unit{rad^2/Hz}, as displayed by the PN analyzer.  The same spectra, corrected for the gain, are shown Figure~\ref{fig:Results-detailed}\,D\@.  This is the AM noise of the two channels of the AD9144.  The unit is \unit{1/Hz}, as appropriate.  As before, the results overlap on almost the full span, but for a small discrepancy in the upper half decade.  Here, the noise measured with the lower gain (20 dB) is some 1.5 dB higher.  The bump at 100 kHz, 5 dB above the asymptotic approximation, is probably due to the power supply.  We exclude the AD9144 internal reference because it is a common-mode signal, and has at most second-order effect on the amplitude noise.

In the second experiment, we measure amplitude and phase noise using the AN detector (Figure~\ref{fig:Results-detailed}\,E and \ref{fig:Results-detailed}\,F).  The maximum frequency is limited by the full span of the FFT analyzer, which is of 100 kHz.  The AM noise spectra overlap well on most of the span (Fig.~\ref{fig:Results-detailed}\,E).  A small spread, $\pm1$~dB, shows up in the upper decade.  The almost-flat region beyond 10 kHz is actually the bump already seen in the AM noise spectra (Figure 7D).  The PM noise spectra (Fig.~\ref{fig:Results-detailed}\,F) overlap well, with a maximum spread of $\pm 1.5$ dB in the upper decade.  In both cases, the higher noise is observed with the lowest gain, 20 dB.  This is ascribed to the background noise of the AM detector, which cannot be rejected. 

Because a white floor is not visible in any plot of Fig.~\ref{fig:Results-detailed}, we can only evaluate the upper bound and check on the consistency with the design parameters.  From the value $S_\alpha=-161$ \unit{dB/Hz} based on the data sheet, and accounting for 1 dB contribution of the RF amplifier, we expect $S_\alpha=-160$ \unit{dB/Hz} for the two converters.  This is exactly equal to the lowest value seen on Fig.~\ref{fig:Results-detailed}\,D  (40 \unit{dB/rad} and 49 \unit{dB/rad} gain) at $f=850$ kHz.  
Inspecting on $S_\varphi$, the lowest value seen on Fig.~\ref{fig:Results-detailed}\,D  (40 \unit{dB/rad} and 49 \unit{dB/rad}) is of $-159$ \unit{dB/Hz}  at $f=850$ kHz, that is, 1 dB higher than the AN\@.  This indicates that the measured values are consistent with the design.

\begin{figure}[t]\centering
\includegraphics[scale=0.459]{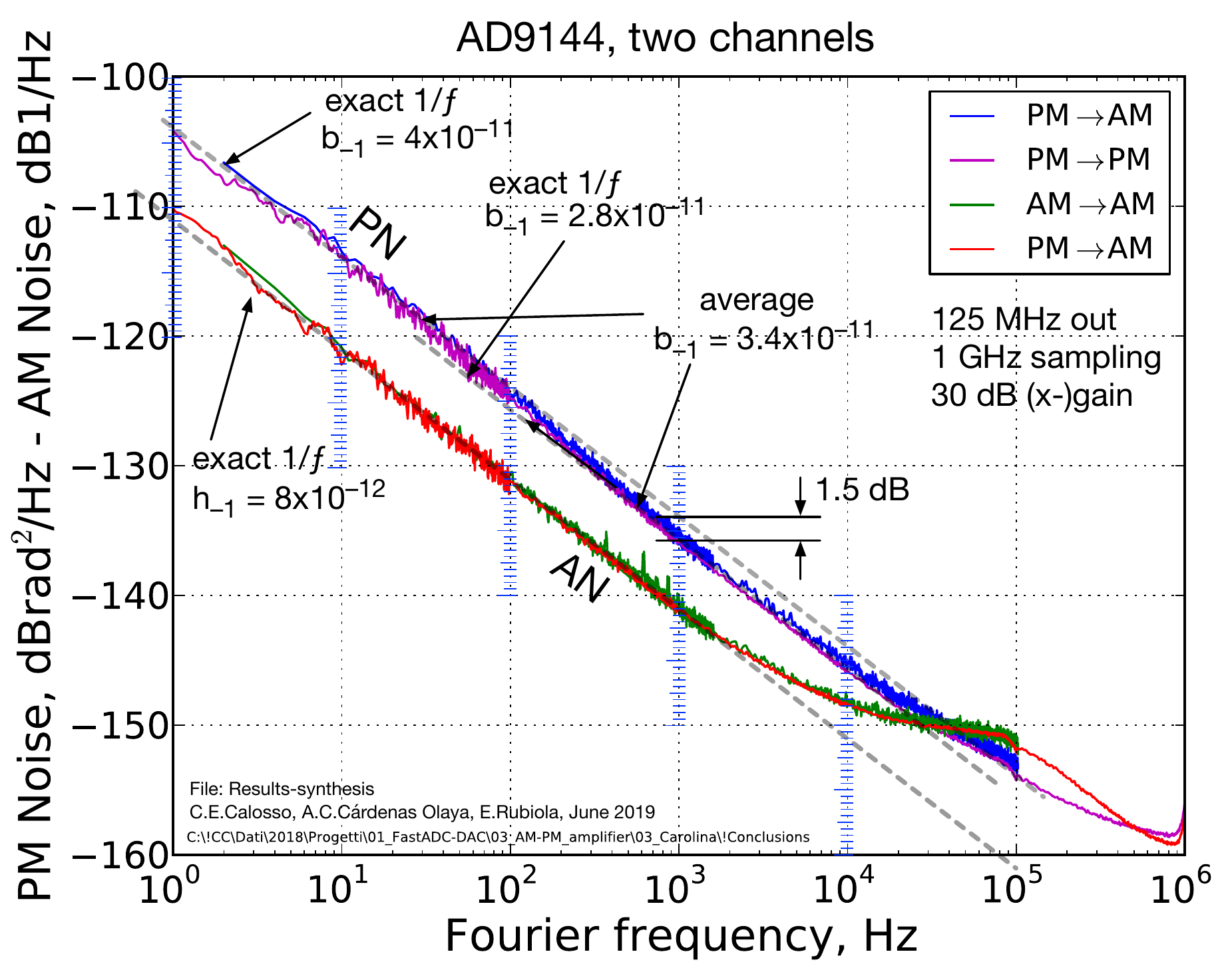}
\caption{Comparison of the results with 30 dB gain.  The full span of the FFT analyzer limits the \unit{AM\,\rightarrow\, AM} and \unit{PM\,\rightarrow\,AM} spectra to 100 kHz.}
\label{fig:Results-synthesis}
\end{figure}

Figure~\ref{fig:Results-synthesis} compares the above results with 30 dB gain.  The two AM noise spectra overlap, and likewise the two PM noise spectra.   This confirms that the two variants of the method, AN analyzer and PN analyzer, give equivalent results.  

The flicker of amplitude is $\mathsf{h}_{-1}=8{\times}10^{-12}$ ($-110$ dB/Hz at 1 Hz) up to a few kHz, corrupted by a bump at 100 kHz.  Such flicker is equivalent to a fractional amplitude stability $\sigma_\alpha=3.3{\times}10^{-6}$ (Allan deviation).   This is found using the classical formula $\sigma^2=2\ln(2)\,\mathsf{h}_{-1}$, which holds for flicker \cite{IEEE-STD-1139-2008,Calosso-2016-Digital}.  

The flicker of phase takes two levels, $\mathsf{b}_1=4{\times}10^{-11}$ \unit{rad^2} ($-104$ \unit{dBrad^2}) at lower $f$ and $\mathsf{b}_1=2.8{\times}10^{-11}$ ($-105.5$ \unit{dBrad^2}) at higher $f$, with a discrepancy of 1.5 dB.  The average of these two values is $\mathsf{b}_1=3.4{\times}10^{-11}$ \unit{rad^2}.  The latter, converted into time-fluctuation PSD is  $S_\mathsf{x}=\mathsf{k}_{-1}/f$ with $\mathsf{k}_{-1}=5.5{\times}10^{-29}$ \unit{s^2}.  Using $\sigma^2=2\ln(2)\,\mathsf{h}_{-1}$, we find a time fluctuation $\sigma_\mathsf{x}=8.7$ fs.  
Having tested only one frequency, we are still unable to divide the \emph{time type noise} (the time fluctuation $\sigma_\mathsf{x}$ is independent of the carrier frequency $\nu$) and the \emph{phase type noise} (the phase fluctuation in rad is independent of $\nu$, and $\sigma_\mathsf{x}$ scales as $1/\nu$) that combine into this result (see \cite{Calosso-2016-Digital} for definitions and properties of phase-type and time-type noise). 

\begin{figure}\centering
\includegraphics[scale=0.459]{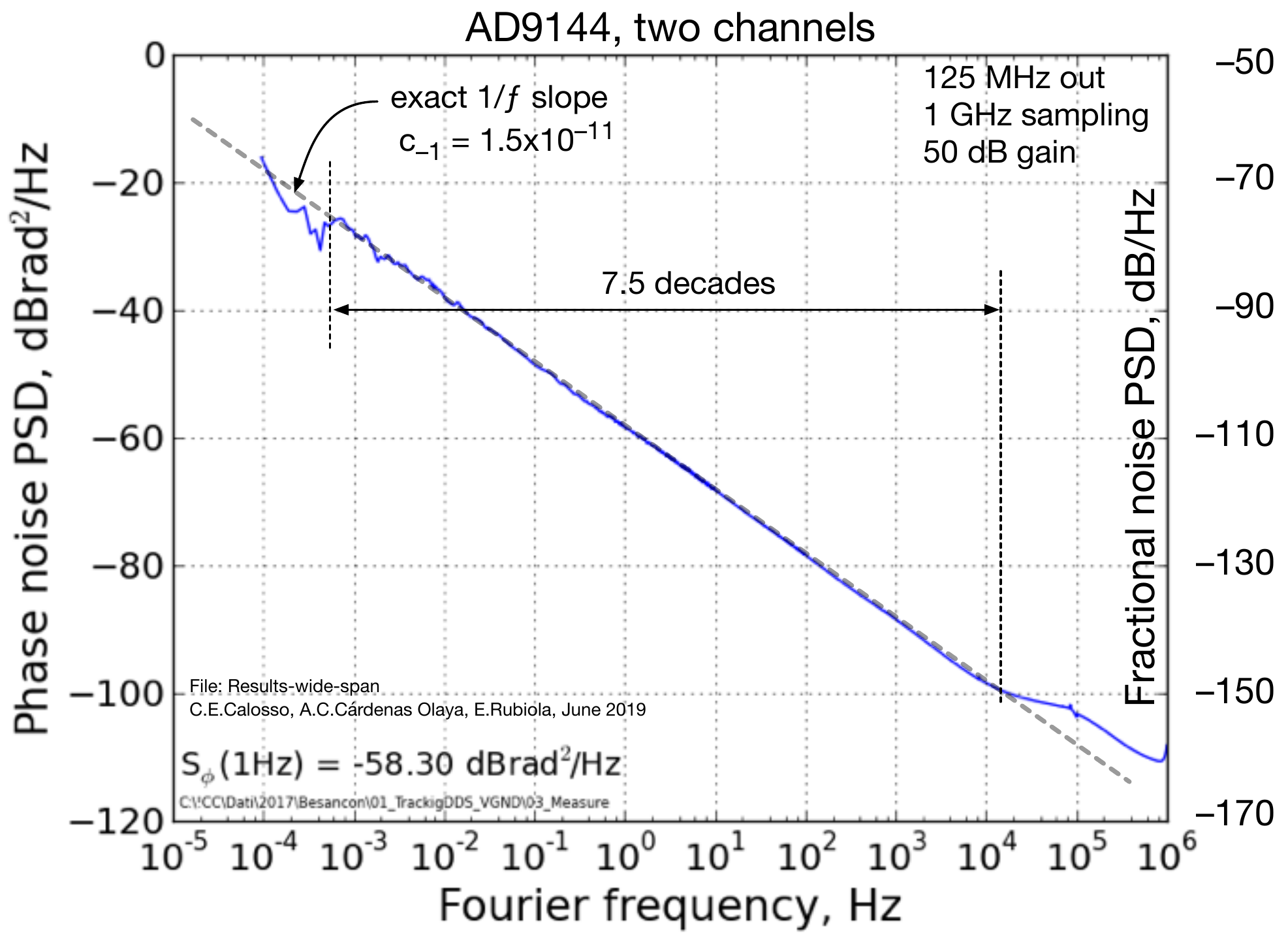}
\caption{Flicker noise observed on the widest span.  Because $\theta $ is not calibrated, the measured quantity (right-hand scale) is a combination of AM and PM.}
\label{fig:Results-wide-span}
\end{figure}

Figure~\ref{fig:Results-wide-span} shows the $1/f$ noise measured on the widest span.  The experiment was done in Besan\c{c}on with the alternate configuration mentioned.  The calibration is unfortunately less reliable than in the other cases.  The gain is of 50 dB, but the phase of the residual carrier was not correctly set.  Thus, we assessed $\theta$ a posteriori using the flicker coefficients of 
\begin{equation}
S_\psi = \frac{1}{{\eta }^2}
\Bigl[\sin^2(\theta)\,S_\alpha + \cos^2(\theta)\,S_\varphi\Bigr]~.
\end{equation}
The observed quantity is $\eta^2S_\psi=\mathsf{c}_{-1}/f$ with $\mathsf{c}_{-1}=1.5{\times}10^{-11}$ (from Fig.~\ref{fig:Results-wide-span}), and the reference quantities are $\mathsf{h}_{-1}=8{\times}10^{-12}$ and $\mathsf{b}_{-1}=3.4{\times}10^{-11}$ (average of the two levels shown), found on Figure~\ref{fig:Results-synthesis}.  Solving
\begin{equation}
\mathsf{c}_1=h_{-1}\left(1-\cos2\theta\right)+\mathsf{b}_{-1}\cos^2\theta~,
\end{equation}
we find $\cos^2\theta=0.26$, and finally $\theta=1.03$ rad ($59^\circ$).  Accordingly, the observed result is 
\begin{equation}
\eta^2 S_\psi=\left[0.74\,S_\alpha + 0.26\,S_\varphi\right]~.
\end{equation}
Regardless of the accuracy of $\theta$, the relevance of this result is the observation of the flicker noise with exact $1/f$ slope over 7.5 decades, with a maximum discrepancy of 1 dB\@.

\section{New Perspectives}
\begin{figure}[t]\centering
\includegraphics[width=\textwidth]{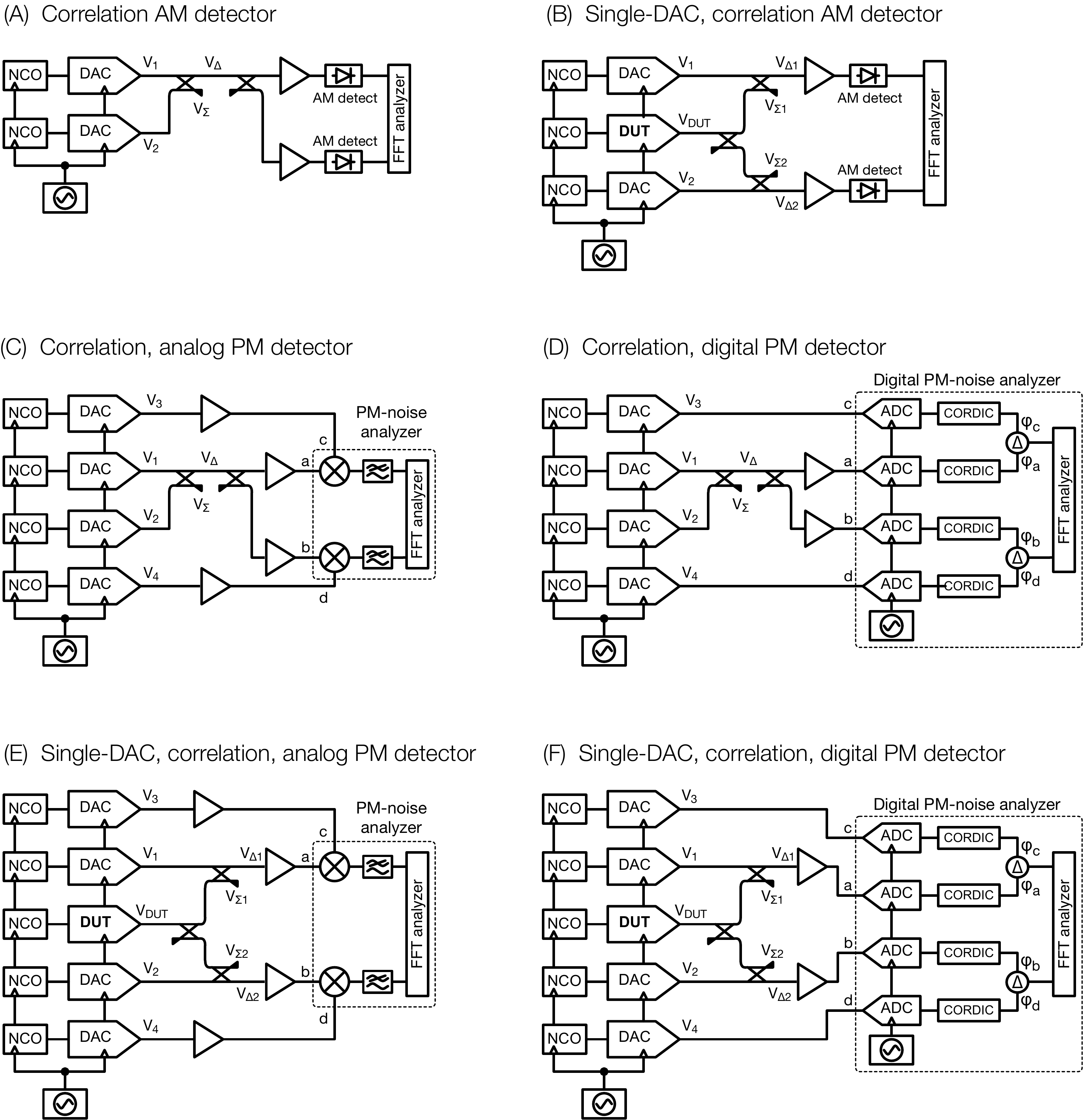}
\caption{New schemes, under consideration for the next experiments.}
\label{fig:New-ideas}
\end{figure}

The lesson learned, combined with our past experience \cite{Rubiola-2005-AM,Rubiola-2002-RSI}, suggests the schemes shown on Figure~\ref{fig:New-ideas}.  These schemes are neither peer-reviewed nor tested, but they rely on solid experience.  Two ideas are used in different configurations, yet for the same purpose.  

The first idea (Fig.~\ref{fig:New-ideas}\,A-C-D)  is that $V_\Delta$ can be split into two branches, amplified, and sent to the two inputs of a dual-channel noise analyzer.  The noise of the two branches is statistically independent, and rejected by correlation and averaging.

The second idea (Fig.~\ref{fig:New-ideas}\,B-E-F) is to use two reference DACs (`1' and `2') to generate two separate `$V_\Delta$.'  The correlation-and-averaging  process extracts the statistical characteristics of $V_\text{DUT}$, and rejects the noise of $V_1$ and $V_2$.

The configuration of Fig.~\ref{fig:New-ideas}\,A is the straightforward application of the first idea.  After rejecting the noise of the RF amplifier and of the AM detector, the cross-spectrum converges to the total noise PSD of the two DACs.

Figure~\ref{fig:New-ideas}\,B enables the measurement of a single DAC by correlating two independent branches, each of which amplifies the modulation index.  The noise of the two reference DACs ($V_1$ and $V_2$) is independent, and only $V_\text{DUT}$ is common to the two branches.

Some PN analyzers provide access to all the four inputs.  This is exploited in Fig.~\ref{fig:New-ideas}\,C-F\@.  In the case of the analog instruments, two amplifiers are needed to increase $V_3$ and $V_4$ to the saturation level of the mixer.
The Symmetricom instruments \cite{Microsemi-ADEV} have only two accessible inputs, thus they are not suitable to these schemes.  By contrast, the forthcoming  PhaseStation \cite{PhaseStation} has all the four ADC inputs accessible.  

Figure~\ref{fig:New-ideas}\,C is the obvious remake of Fig~\ref{fig:New-ideas}\,A, after replacing the AM detector with a PM detector.
The two additional DACs, `3' and `4,' provide the phase reference to the PN analyzer.  One more adjustment is needed, to set $V_3$ and $V_4$ in quadrature to the amplified version of $V_\Delta$.

Figure~\ref{fig:New-ideas}\,D is equivalent to Fig.~\ref{fig:New-ideas}\,C, but for the digital analyzer instead of the classical saturated-mixer instrument.  In this case there is no need to set the phase of $V_3$ and $V_4$.  Trusting the modulation-index amplification, it is possible to save one converter (`4') feeding the phase references `c' and `d' with the same signal ($V_3$).  The reason is that the digital instrument accept arbitrary phase (and frequency) at the reference inputs.

Figure~\ref{fig:New-ideas}\,E is the obvious remake of Fig~\ref{fig:New-ideas}\,B, replacing the AM detector with a PM detector.  Because only $V_\text{DUT}$ is common to the two branches, this configuration enables the measurement of a single DAC by correlating the two independent branches, each of which amplifies the modulation index.
One may be tempted to save two DACs by feeding the phase references `c' and `d' with $V_{\Sigma1}$ and $V_{\Sigma2}$.  This is probably not a good idea.  A problem is that the phase of such signals cannot be controlled independently of $V_\Delta$, thus the quadrature condition has to be set by other means.  Another problem is that even a small change in the impedance seen by $V_\Sigma$ impacts dramatically on $V_\Delta$, thus the gain $1/\eta$ and on the phase $\theta$.

Figure~\ref{fig:New-ideas}\,F is equivalent to Fig.~\ref{fig:New-ideas}\,E, but for the classical analog instrument replaced with a fully digital PN analyzer.  As before, it is possible to save one converter (`4') feeding the phase references `c' and `d' with the same signal ($V_3$), but the signals `$V_\Sigma$' should not be re-used.

\section{Conclusions}

We have proposed a method for the measurement of PM noise AM noise of DACs and DDSs, and we have proved the concept by measuring an AD9144.

The value of the method is in its reliability and simplicity.  Implementation and use require modest skill in analog RF electronics, and only standard skill in programming and using the target DACs and DDSs.  Our experiments rely on commercial parts only, like the Z-Board and DAC daughterboards, and on ready-to-use connectorized RF modules.  Under no circumstance we had to design and implement ad hoc electronics.  The low background noise is inherent in the principle, and easy to achieve.  The cross spectrum comes, optionally, only after modulation-index amplification.  Consequently, none of the known flaws of the cross spectrum can threat the reliability of the result.  Tuning and calibration can be automated because it is done entirely by setting integer numbers in the target converter.  Thanks to all these characteristics, our method has the potential to become the standard method for the measurement of AM noise and PM noise of DACs and DDSs.

The experimentalist has two options, using a dedicated PN analyzer or a simple power detector and a general-purpose FFT analyzer.  Complexity and background noise are equivalent, and the results overlap. 

The power-detector option does not require a PN analyzer, which is a specialized and expensive instrument.  The FFT analyzer can be implemented in GNU radio on the Z-Board, after adding an ADC daughterboard.  Overall, this version probably fits in the budget of a radio amateur or hobbyist.

The other option requires a commercial PN analyzer.  In fact, the design and implementation of such instrument is definitely not simple.  The virtue of a fully-digital PN is the wide dynamic range, inherent in the CORDIC algorithm used to extract the random phase modulation from the sampled input \cite{Meher-2009}.

As a fringe benefit, we observed the noise of the AD9144 on a frequency span of 10 decades.  The flicker noise matches the exact $1/f$ law, with a maximum discrepancy of $\pm 1$ dB over 7.5 decades.  The flicker noise of an electronic device over such a wide frequency range has probably never been reported before.

\section*{Acknowledgments}
This work is funded by the ANR Programme d'Inves\-tis\-se\-ment d'Avenir under the Oscillator IMP project (ANR-11-EQPX-0033-OSC-IMP) and the First-TF network (ANR-10-LABX-48-01), and by the Région Bourgogne Franche Comté.

\bibliography{Ref-short,Ref-local,Ref-Rubiola,References}

\end{document}